# Intelligent Mode-Locked Single-Cavity Dual-Comb Laser Utilizing Time-Stretch Dispersive Fourier Transform Spectroscopy

PAN GUO,[1] YUAN GAO,[1] YONGJIE PU,[1] ZHIGANG ZHAO,[2] ZHENHUA CONG,[2] AND SHA WANG[1,*]

[1]*College of Electronics and Information Engineering, Sichuan University, 610064, Chengdu, China*
[2]*School of Information Science and Engineering, and Shandong Provincial Key Laboratory of Laser Technology and Application, Shandong University, 266237, Qingdao, China.*
**Correspondence: shawang@scu.edu.cn*

**As single-cavity dual combs play a significant role in numerous high-precision measurements, their efficient generation has been widely researched. To achieve stable dual-comb mode locking remains, a two-part evaluation criterion containing a fitness function and a CNN-Transformer network is employed. Simulated time-stretch dispersive Fourier transform (DFT) spectra are used as datasets, which simplifies the optimization process and does not rely on specific experimental data. A developed evolutionary algorithm (EA) for paddle-based motorized polarization controllers (MPCs) is proposed, enabling the intelligent attainment of dual-comb mode-locked states. A real-time library stores fitness and MPC angles, facilitating mode-locked state achievement within 2 seconds. Finally, long term running of dual-comb mode locking is ensured by a random collision algorithm utilizing an evaluation criterion of weak soliton peaks.**

Dual-comb technologies involve utilizing two coherent optical pulse trains with a small frequency difference in their repetition rates, which can offer capabilities for precise ranging [1-2] and molecular spectroscopy [3-4]. Compared to combining two independent optical combs, employing a single-cavity dual-comb laser eliminates the need for additional stabilization approaches to lock the phases of the two pulse trains. And a high mutual coherence is achieved due to common-mode noise cancellation [5].

As the generation of single-cavity dual combs fundamentally rely on the concept of "multiplexing", the intricate interplay of intracavity dispersion, nonlinear dynamics, and environmental perturbations collectively induces instability in the single-cavity dual-comb formation. For that, intelligent mode-locked techniques based on artificial intelligence and electric polarization controllers have been developed [6-9]. However, fewer research focuses on intelligent dual-wavelength mode-locking systems. The intricate interplay of intracavity dispersion, nonlinear dynamics, and environmental perturbations collectively induces instability in the dual-comb formation. Thus, establishing and maintaining the dual-comb mode locking within a single cavity remains a challenge, which restricts the practical utility of dual combs. In 2023, G. Pu et al. demonstrated the first intelligent single-cavity dual-comb fiber laser by combining the real-time intelligent control and the memory-aided intelligent searching (MAIS) algorithm. Employing a field-programmable gate array (FPGA), and two digital-to-analog converters enabled MAIS can locate a desired solution in a mean time of only 2.48 seconds [10]. Q. Yan et al. also reported an automatic mode locking in a dual-wavelength soliton fiber laser. An oscilloscope and an optical spectrum analyzer were used simultaneously to record temporal pulse trains and spectral information. By using a two-stage genetic algorithm, the laser kept a stable dual-wavelength mode locking for hours [11]. The time taken to implement dual-wavelength mode locking has not been mentioned.

In this work, we report an intelligent single-cavity dual-comb mode-locked fiber laser based on time-stretch dispersive Fourier transform (DFT). An oscilloscope is used to receive DFT spectra in real time, which combines time-domain and spectral information. A two-part evaluation criterion containing a fitness function and a CNN-Transformer network is utilized. It ensures that both global and local characteristics of the time-stretch spectrum are comprehensively considered. The utilized datasets exclusively consist of synthetically generated time-stretch DFT spectra, that can leave out the early collection of empirical data and improve the universality of the model. A developed EA is proposed to achieve intelligent mode locking of the dual-comb state. High-order harmonic mode locking can also be accurately distinguished according to their spectral information. A real-time updatable library is constructed, encompassing diverse mode-locked states. To ensure the reliability of the stored states, a state filtering is implemented. Experimental results demonstrate that the mode-locked library can be established within a 15 minutes. The saved angles in the library can be directly used to achieve the corresponding dual comb mode-locked states, with an average time of less than 2 seconds, which is the fastest record according to our best knowledge.

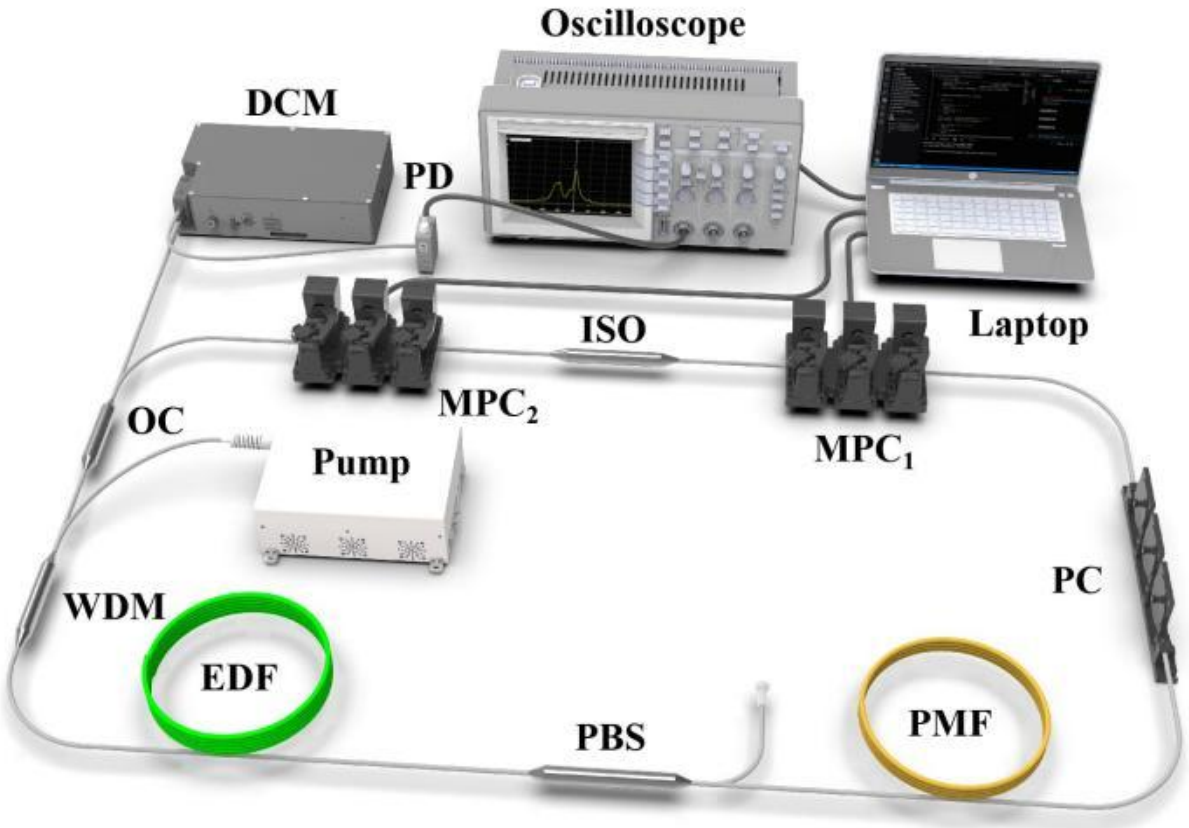

Fig. 1. The experimental setup of the intelligent single-cavity dual-comb mode-locked laser.

---

The configuration of the intelligent dual-comb fiber laser is shown in Figure 1. A 2.5-m polarization-maintaining erbium-doped fiber (PM-EDF) with a calculated second-order dispersion of 28 $ps^2$/km constitutes the gain medium, pumped by a 980-nm laser diode through a 980/1550 wavelength division multiplexer (WDM). A polarization-dependent isolator (PD-ISO) ensures the unidirectional transmission of light in the fiber, which together with two MPCs facilitates the nonlinear polarization rotation (NPR) mode locking. A separate PBS-PMF-PC structure is adopted to establish filtering effect. The total length of the laser cavity is ~17 m. The net cavity dispersion is ~-0.308 $ps^2$ at 1550 nm which means the mode-locked laser operates in the anomalous regime. Approximately 90% of the laser energy stays within the ring cavity by using an output coupler (OC). The output laser is propagated through a dispersion compensation module (DCM) with a dispersion of DL = -654 ps/nm. A laptop that runs EA and is connected to the oscilloscope can receive time-stretch DFT spectra in real time through a balance photodetector (PD). And then it sends new control commands generated by the EA to electrically control six paddles of the two MPCs directly to achieve polarization control.

Establishing an excellent evaluation criterion is the key premise of intelligent search. The evaluation criterion of the individual's competitiveness has two parts: fitness function and neural network classification. The dual-region count scheme is adopted to judge the pulse trains preliminary [12]. Two intensity thresholds $T_1$ and $T_2$ are set, where $T_1 > T_2$. $T_1$ is built for the desired pulse count. $T_2$ is a limitation that noise should not surpassed. Specifically, the DFT spectrum on the oscilloscope greater than $T_1$ is considered as effective pulses, and those less than $T_2$ are considered as noises. The fitness function can be constructed as follows:

$$Fitness = I_{pluse} - \alpha \left| C_{real} - C_{ideal} \right| - \beta I_{noise} \quad (1)$$

Where the first term of the fitness is the sum intensity of the pulse $I_{pluse}$ within a fixed time window; The second term represents the deviation punishment for the pulse count with a scale factor $\alpha$, where $C_{real}$ and $C_{ideal}$ denote the real and ideal pulse count, respectively; The third term stands for the punishment of noise $I_{noise}$ and its weight $\beta$. Five fundamental mode-locked pulse can be observed completely. Consequently, when the value of $C_{ideal}$ is set to 5 and 10, Equation (1) denotes the single-wavelength mode-locked fitness and the dual-comb mode-locked fitness, respectively. After balancing the influence of all components in the fitness function, $\alpha$ and $\beta$ are set to 0.2 and 0.01, respectively.

Moreover, the pulse trains of the second-order harmonic and dual-comb mode locking are very similar, this makes it difficult to efficiently distinguish with the fitness function. Therefore, the second evaluation of the individual's competitiveness, a CNN-Transformer network, is proposed to classify the one-dimensional time-stretch DFT spectra in time domain. The training datasets of the CNN-Transformer network are completely provided by numerical simulations of the Ginzburg-Landau equation.

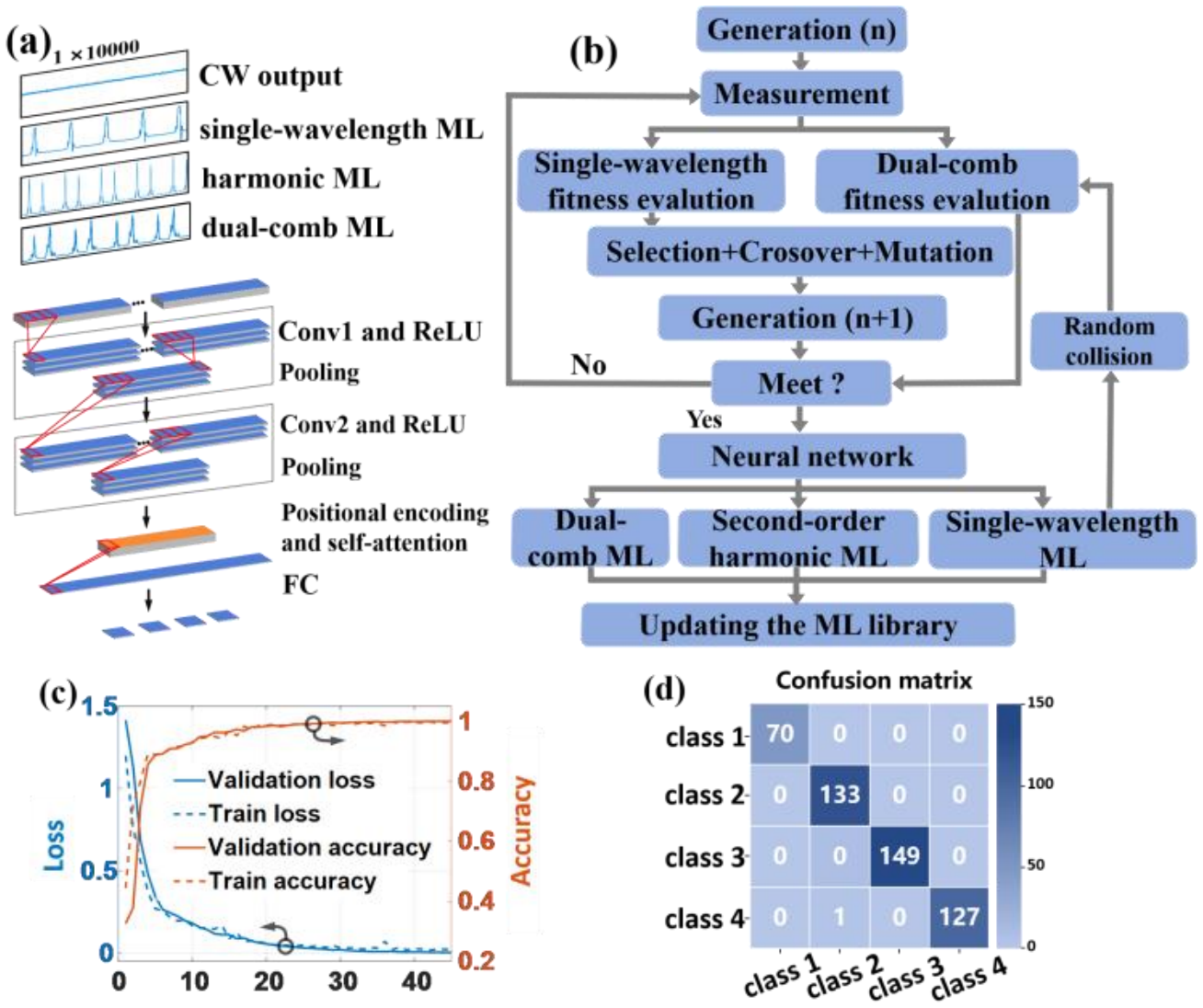

Fig. 2. (a) The structure of the CNN-Transformer network. (b) The principle of the developed EA. (c) The curves of the loss and accuracy over epochs. (d) Confusion matrix of 4 classifications. Class 1 through 4 correspond to continuous wave (CW), single-wavelength, harmonic, and dual-comb mode locking, respectively.

---

The network structure is shown in Figure 2(a). Convolutional layers in CNN are excellent at extracting local features (e.g., spikes or small variations) in pulse trains, while the positional encoding and self-attention in Transformer can extract the global features. The DFT spectra are divided into the training, validation, and test sets in a 7:1.5:1.5 ratio. The accuracy and loss results of the train and validation are demonstrated in Figure 2(c). The accuracy of the validation sets of the model reaches 99.79% within 50 epochs, while the accuracy of the training sets is 100%. From the confusion matrix depicted in Figure 2(d), it can be observed that the periodic overlap resulting from temporal asynchrony occasionally presents a challenge to its correct classification. In addition, we classify the experimental DFT spectra with this trained network. Benefit from the robust and property data processing capabilities of the CNN-Transformer network, the accuracy is close to 100%. Thus, the second-order harmonic can also be accurately distinguished. During the entire optimization process, laser outputs corresponding to different mode-locked states are not collected in advance. Even though higher-order harmonic mode locking above the third order is not observed due to the limitation of pump power, it is still possible to quickly search for high-order harmonic mode locking based on the spectral information contained in the time-stretch DFT pulses, as well as by properly setting the ideal pulse count of the target mode-locked state Cideal in the fitness function of Equation (1). Therefore, this approach not only does not require experimental prior data but also enhances the universality of the model across various fiber laser systems.

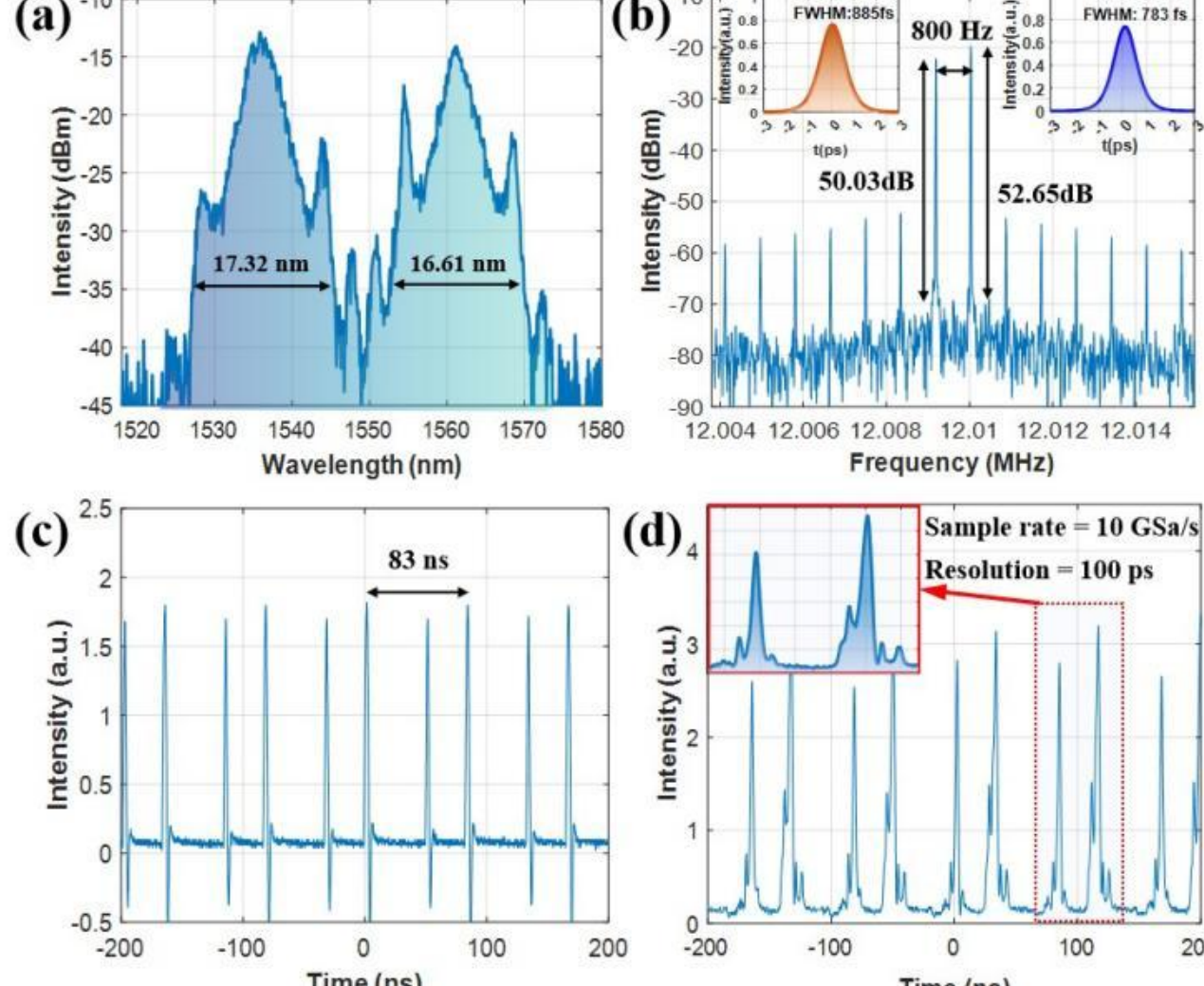

Fig. 3. (a) Optical spectrum. (b) RF spectrum with beat frequency, insets: autocorrelation trace of solitons centered at 1536 nm and 1561 nm, respectively. (c) Time-domain pulse trains. (d) Time-stretch DFT spectrum, inset: local view of a peak in a single period.

---

The intelligent searching of dual combs is achieved via a developed EA, as shown in Figure 2(b). In this proposed algorithm, each individual has six genes, associated with the six control voltages of two MPCs. Then the user-defined fitness functions of each existing individual are calculated by collecting the dispersed consecutive waveform of time-stretch pulses through the oscilloscope. In general, single-wavelength fitness is used for achieving automatic single-wavelength mode locking. After that, a random collision algorithm that aims to realize the dual-comb mode locking. Random rotation angles with Gaussian probability distribution between -3° and 3° are selected in the process of random collision. This angle selection method can improve the search speed while ensuring search accuracy. We also set a threshold of 20 times random collisions. When the dual-comb mode locking is not found after the threshold number is exceeded, the EA will continue to iterate to find the next single-wavelength mode locking.

To prevent repetitive searching of mode locking in the same laser, we build a library containing the fitness function for different stable mode-locked states and the six angles of MPCs. The data in the library supports real-time updating of new mode-locked states and reading of all states. State filtering is added to ensure the reliability of the state in the library. After the completion of the search of each generation in the evolutionary algorithm, angles of the six paddles corresponding to the mode-locked state stored in the library will be read for inspection. If the state is different from that stored, it will be removed from the library and continue the next round of search. Depending on the initial angles of MPCs' paddles, after multiple experiments, it is verified that the library can be built up within 15 minutes. After the mode-locked library loses effectiveness, the library can also be rebuilt with EA running. Reading the saved angles can directly achieve the corresponding mode-locked state, with an average time of less than 2 seconds.

As shown in Figure 3(a), the spectral widths of separated mode-locked pulses with central wavelengths of 1535.83 nm and 1561.26 nm are 17.32 nm and 16.61 nm, respectively. Self-phase modulation (SPM) is influenced by the intracavity nonlinear coefficient and pulse intensity, will alter the pulse phase, thereby leading to a broader spectrum at 1536-nm mode locking. The total output power is about 2 mW under a pump power of 220 mW. The RF spectrum can be seen in Figure 3(b), two repetition-rate signals can be found around 12.0 MHz, and the corresponding repetition-rate difference is about 800 Hz. In addition, the signal-to-noise ratios of both repetition rates are all over 50 dB, and the linewidth of the beat RF signal is about 2.34 Hz with the 1 Hz resolution bandwidth. The insets present the second-harmonic generation autocorrelation trace of output solitons centered at 1536 nm and 1561 nm, respectively. Their pulse duration measured by autocorrelator is about 885 fs and 783 fs, respectively. The time-bandwidth products for the 1536 nm and 1561 nm pulses are 1.9491 and 1.6012, respectively. The employment of the Lyot filter allows for the realization of single soliton mode locking at a higher pump power [13], so the phenomenon of multiple solitons caused by pulse splitting is not observed. The temporal oscilloscope waveform of the dual-comb pluses is shown in Figure 3(c), with a measured pulse period of ~ 83 ns. Figure 3(d) displays the spectra after time-stretch DFT, the inset presents the local view within a single time period.

Compared with single-wavelength mode locking, dual-comb mode-locked states, whose performance is highly dependent on the delicate balance between intracavity dispersion and birefringence, is more difficult to maintain stable operation. We propose a stability maintenance algorithm for dual-comb mode locking using random collision. The spectrum changes during the stable operation of the dual-comb mode-locked state for over 6 hours are shown in Figure 4(a). The evaluation and specific processes of this algorithm are detailed as follows. In the dual-comb mode-locked states, two solitons with different center spectra will travel with different group velocities. Therefore, when one of the solitons is triggered, another one will periodically move on the oscilloscope. It is observed in the experiment that the triggered soliton is usually the dominant one while the moving soliton is usually weaker. This is related to the balance between the intracavity gain and loss in the laser and the setting of the Lyot filter. The stable operation of weak solitons is the key to maintaining the dual-comb state. The standard deviations of several weak solitons peaks are calculated as an evaluation criterion in real time. As shown in Figure 4(b), changes of the standard deviation in the weak soliton peaks are recorded within 6 hours, and the inset depicts a detailed example of weak soliton peak selection. Once the standard deviation is observed to exceed the threshold of 0.08, the stability maintenance algorithm with random collision is activated. After each random adjustment, the standard deviation of the output signal is recalculated to assess the effect of the collision. If the standard deviation remains above the threshold, the system counteracts the previous rotation by reversing the paddle to its original position. Subsequently, a new random collision adjustment is initiated, and the process iterates. This approach ensures a robust restoration of the dual-comb state while maintaining precision, as the iterative process balances randomness with reversibility.

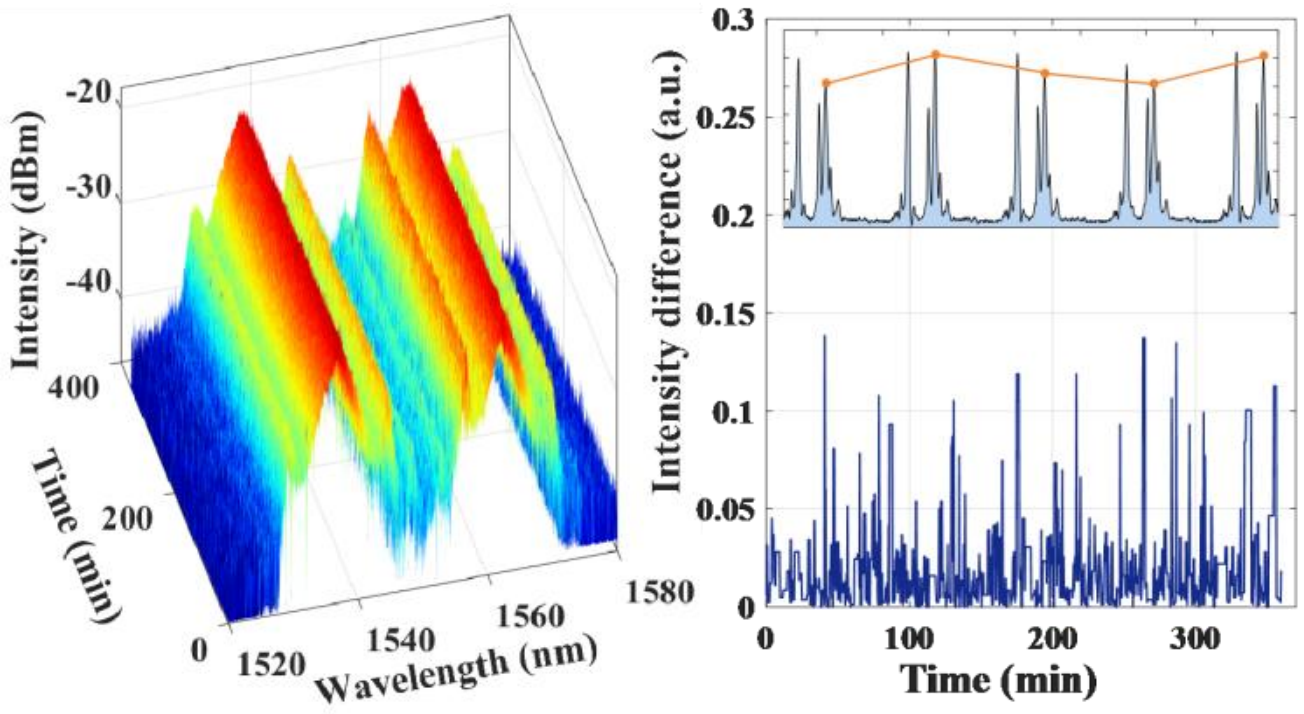

Fig. 4. (a) Stability maintenance of dual combs. (b) Changes of the standard deviation of pulse peak values in the process of stability maintenance, inset: details of the soliton peak selection.

---

In conclusion, we demonstrate an intelligent single-cavity dual-comb mode-locked fiber laser. A two-part evaluation criterion containing a fitness function and a CNN-Transformer network is proposed to elevate dual-comb mode-locked states. The input datasets corresponding to various mode-locked states are obtained by simulation, which circumvents the need for over-reliance on specific experimental datasets. Then a developed EA is utilized to optimize the fitness function and control paddle-based MPCs to find the desired mode-locked state. After that, a library supporting real-time updating and reading is built to store the fitness functions for different stable mode-locked states. The mode-locked library can be set up within a 15-minute duration. Reading the saved angles can directly achieve the corresponding mode-locked state, with an average time of less than 2 seconds. In addition, a new evaluation criterion combined with a random collision algorithm is used to achieve the stable maintenance of the dual-comb mode locking. For subsequent work, we plan to apply it to practical applications such as absorption spectrum measurement and fiber sensing.

**Funding.** This work was supported by the National Natural Science Foundation of China (Grant Nos. 62475175), Opening Foundation of Key Laboratory of Laser & Infrared System (Shandong University), Ministry of Education.

**Disclosures.** The authors declare no conflicts of interest.

**Data availability.** Data underlying the results presented in this paper are not publicly available at this time but may be obtained from the authors upon reasonable request.